\documentclass[]{spie}  
\usepackage[]{graphicx}
\usepackage{natbib}

\title{
Beating the confusion limit:
The necessity of high angular resolution for probing the
physics of Sagittarius A* and its environment:
\\
Opportunities for LINC-NIRVANA (LBT), GRAVITY (VLTI) and and METIS (E-ELT)
} 

\author{A. Eckart\supit{a,b},
N. Sabha\supit{a},
G. Witzel\supit{a,c},
C. Straubmeier\supit{a},
B.~Shahzamanian\supit{a,b},
M. Valencia-S.\supit{a,b}
Macarena Garc\'{\i}a-Mar\'{\i}n\supit{a},
M. Horrobin\supit{a},
L. Moser\supit{a},
J. Zuther\supit{a},
S. Fischer\supit{a},
C. Rauch\supit{a},
S. Rost\supit{a},
C. Iserlohe\supit{a},
S. Yazici\supit{a},
S. Smajic\supit{a},
M. Wiest\supit{a},
C. Araujo-Hauck\supit{a}  and
I. Wank\supit{a}
\skiplinehalf
\supit{a}University of Cologne, I. Physikalisches Institut, 
Z\"ulpicher Str. 77, 50937 K\"oln, Germany; \\
\supit{b} MPIfR, Auf dem H\"ugel 69, Bonn, Germany
\supit{c} UCLA, Los Angels, USA
}

\authorinfo{Send correspondence to A.Eckart, 
E-mail: eckart@ph1.uni-koeln.de, Telephone: +49 221 470 3546}

\newcommand{\solm}{M$_{\odot}$\ }

\newcommand{\rf}{\par\noindent\hangindent 15pt {}}

\begin{document} 
  \maketitle 

\begin{abstract}
The super-massive 4 million solar mass black hole (SMBH) SgrA* shows 
variable emission from the millimeter to the X-ray domain. 
A detailed analysis of the infrared light curves allows us to 
address the accretion phenomenon in a statistical way. 
The analysis shows that the near-infrared flux density excursions
are dominated by a single state power law, with the low states 
of SgrA* limited by confusion through the unresolved stellar background. 
We show that for 8-10m class telescopes blending effects 
along the line of sight will result in artificial compact star-like 
objects of 0.5-1 mJy that last for about 3-4 years. 
We discuss how the imaging capabilities of GRAVITY at the VLTI, 
LINC-NIRVANA at the LBT and METIS at the E-ELT will 
contribute to the investigation of the low variability states of SgrA*. 	 
\end{abstract}


\section{INTRODUCTION}
\label{sec:intro}  

The thorough investigation of stellar number densities, luminosities and orbits
as well as light curves of SgrA* allows a detailed analysis of the immediate
surroundings of the super massive black hole at the center of the Milky Way.
However, these investigations also have shown that there are limits for the currently
employed methods. 
Determining the K-band luminosity function (KLF) of the S-star 
cluster members from infrared imaging and  
using  the shape of the stellar distribution 
from number density counts, 
Sabha et al. (2012) have been able to shed some light on the 
amount and nature of the stellar 
and dark mass that is or may be associated with the cluster 
of high velocity S-star cluster in the immediate vicinity of Sgr~A*.
However, for both the identification of faint stars and the investigation
of faint flux density levels of SgrA* the stellar confusion limit as present for
8-10m class telescopes imposes some of these limits. 

In addition to problems simply arising from the stellar confusion limit,
there are other limitations that arise from aliasing populations of massive objects
or disturbing emission mechanisms or confusing emission from particular source structures.
For completeness we mention these effects here as well.
There will for instance be source components in the extended accretion flow and the
thermal Bremsstrahlung component (SgrA* is located in) that add confusion 
to the determination of the intrinsic variability of SgrA* in the 
radio and X-ray domain, respectively.

It is also expected that
there is a population of dark stellar remnants 
at the bottom of the potential well. 
This is based on dynamical and stellar 
evolutionary arguments (see e.g. Morris et al. 1993) 
with some observational evidence 
through the increase of the number density of X-ray binaries
many of which harbor a stellar black hole as a companion
(Muno et al. 2005).
Using Monte Carlo simulations of two-body relaxation, tidal disruptions 
of stars by Sgr~A* in addition to direct plunges through the event horizon
Freitag et al. (2006) find that 
within 0.01, 0.1 and 1~pc from the Center there are approximately
560, $2.4 \times 10^4$ and $2.1 \times 10^5$~\solm in 10~\solm 
stellar black holes 
and 180, 6500 and $3.4 \times 10^5$~\solm in main-sequence stars.
In addition roughly the same amount of white dwarfs and neutron 
stars are expected.
The interaction between the more massive of these stellar remnants and the
visible high velocities of the central S-star cluster will have an 
influence on the orbital parameters of these sources.

In this article we describe and summarize some aspects of these 
effects and point at the requirements and properties 
of upcoming instrumentation that will, at least partially, help to overcome 
some of these confusion problems. 
Here we concentrate on the role of
LINC-NIRVANA 
(Herbst et al. 2010,
Horrobin et al. 2010)
at the LBT
(Vaitheeswaran et al. 2010),
GRAVITY at the VLTI
(Eisenhauer et al. 2011,
Straubmeier et al. 2010),
and
METIS (Brandl et al. 2010) at the E-ELT
(Gilmozzi, R. \& Spyromilio, J., 2008).
For GRAVITY at the VLTI the positioning (phase-referencing) accuracy will be of the order of
10~$\mu$as at 2$\mu$m wavelength (infrared K$_s$-band). 
The GRAVITY imaging resolution in this band will be 
$\le$4~mas, for LINC-NIRVANA at the LBT the angular resolution will be better than
20~mas in the same band, and  at 3.8~$\mu$m for METIS at the E-ELT the angular
resolution will be about 20~mas as well.

\section{Sources of Confusion towards the Galactic Center} 

\subsection{Number density counts in the central arcsecond} 
\label{sec:number}

Sabha et al. (2012) for the first time investigate the effects of stellar scattering 
in the framework of resonant relaxation for the case of the high
velocity S-stars at the Galactic Center.
They find that if a significant population of 10~\solm black holes is present 
in the stellar mass enclosed by the S2-orbit that amounts 
to a value between $10^3$~\solm and $10^5$~\solm 
(see e.g.  Freitag et al. 2006) then,
for trajectories of S2-like stars, contributions from
stellar scattering will be important compared to the the 
relativistic or Newtonian peri-bothron shifts.
This clearly shows that observing only a
single stellar orbit will by far not be sufficient to
put firm limits on the total amount of 
extended stellar mass and on detecting and establishing the importance 
of relativistic peri-bothron shifts of the orbits.
Sabha et al. (2012) conclude that only by observing a larger number 
of stars (at least a few 10)
will allow to sample the statistics of the effect.
However, if the distribution of 10~\solm black holes is
cuspy then this may become even more difficult towards the center 
and close encounters should become increasingly frequent in this region.

In general, however, the inclusion of scattering 
allows us to simultaneously probe the distribution and composition 
of the extended  mass very close to the SMBH. 
In order to do this the astrometric accuracy must be improved by 
at least
an order of magnitude using either larger single dish telescopes 
or interferometers in the NIR.
Removing the bright cluster members,
Sabha et al. (2012) show that the amount of light from
the fainter S-cluster members is below the amount of residual light 
at the position of the S-star cluster.
While Sabha et al. (2010) estimate that only a
maximum of one third of the diffuse light could
be due to residuals from the PSF subtraction, 
Sabha et al (2012)  find that faint stars at or beyond
the completeness limit reached in the KLF can account 
for only about 15\% of the background light. 
However, it cannot be excluded that an additional amount of 
light may also originate from the accretion
process onto a large number of 10~\solm black holes that
will be present in the central region probed by the S-star cluster.
Higher angular resolution and sensitivity are needed
to resolve the background light and analyze its origin.

Sabha et al. (2010) detected three stars that were either 
previously not known or allowed only a very unsatisfactory 
identification with previously known members of the cluster 
(one of them is S62, as pointed out in Dodds-Eden et al. 2011).
In addition the authors point at the case of the star S3 
which was identified in the K-band in 
the early epochs  1992 (Eckart et al. 1996), 1995 (Ghez et al. 1998) 
and lost after about 3 years in 1996/7 (Ghez et al. 1998,
Genzel et al. 2000).
Sabha et al. (2012) investigate this phenomenon 
by extrapolating the KLF in the central region surrounding Sgr~A*, 
to stars fainter than the faintest source (K=$17.31$) the authors 
detected in their sensitive 30 August and 23 September 2004 dataset.
These dataset showed Sgr~A* to be in a particularly low 
activity state.

\begin{table}[!ht]
\centering
{\begin{small}
\begin{tabular}{ccccc}
\hline
\hline
$K_s$-band       & \multicolumn{3}{c}{Power-law index}   \\
magnitude    &              &              &             \\
cutoff       &  0.19        &  0.30        &  0.35       \\
\hline
\hline
\multicolumn{4}{c}{KLF slope $=0.11$}\\ 
	     &              &              &             \\
20.99        & $0.0510$     & $0.0493$     & $0.0858$    \\
             &              &              &             \\
24.67        & $0.0478$     & $0.0722$     & $0.0765$    \\
\hline
\multicolumn{4}{c}{KLF slope $=0.18$}\\
	     &              &              &             \\
20.99        & $0.1302$     & $0.1103$     & $0.2897$    \\
             &              &              &             \\
24.67        & $0.2224$     & $0.2905$     & $0.4162$    \\
\hline
\multicolumn{4}{c}{KLF slope $=0.25$}\\
	     &              &              &             \\
20.99        & $0.5259$     & $0.6160$     & $0.8987$    \\
             &              &              &             \\
24.67        & $0.8924$     & $0.9444$     & $0.9518$    \\
\hline
             &              &              &             \\
\end{tabular}
\caption{Probabilities of detecting a false star (brighter than $K_s=16.39$) in a $1.28''\times 1.28''$ region.}
\label{montecarlo}
\end{small}}
\end{table}
\begin{table}[!ht]
\centering
{\begin{small}
\begin{tabular}{ccccc}
   &              &              &              \\
   &              &              &              \\
   &              &              &              \\
\hline
\hline
$K_s$-band       & \multicolumn{3}{c}{Power-law index}    \\
magnitude    &              &              &              \\
cutoff       &  0.19        &  0.30        &  0.35        \\
\hline
\hline
\multicolumn{4}{c}{KLF slope $=0.11$}\\
	     &              &              &              \\
20.99        & $0.0250(4)$  & $0.0246(4)$  & $0.0612(5)$  \\
             &              &              &              \\
24.67        & $0.0119(8)$  & $0.0214(10)$  & $0.0271(11)$  \\
\hline
\multicolumn{4}{c}{KLF slope $=0.18$}\\
	     &              &              &              \\
20.99        & $0.0504(6)$  & $0.0522(6)$  & $0.1502(8)$  \\
             &              &              &              \\
24.67        & $0.0491(25)$ & $0.0769(29)$ & $0.1290(35)$ \\
\hline
\multicolumn{4}{c}{KLF slope $=0.25$}\\
	     &              &              &              \\
20.99        & $0.1468(10)$  & $0.2507(12)$  & $0.5111(16)$ \\
             &              &              &              \\
24.67        & $0.3208(98)$ & $0.4761(115)$ & $0.5193(122)$ \\
\hline
             &              &              &             \\
\end{tabular}
\end{small}}
\caption{Probabilities of detecting a false star (brighter than $K_s=16.39$) 
at the position of Sgr~A*. The number of stars contributing to the detected 
flux of the false star is given in parentheses for each considered case.}
\label{montecarlocentral}
\end{table}

\begin{figure*}
  \centering
 \includegraphics[width=17cm,angle=00]{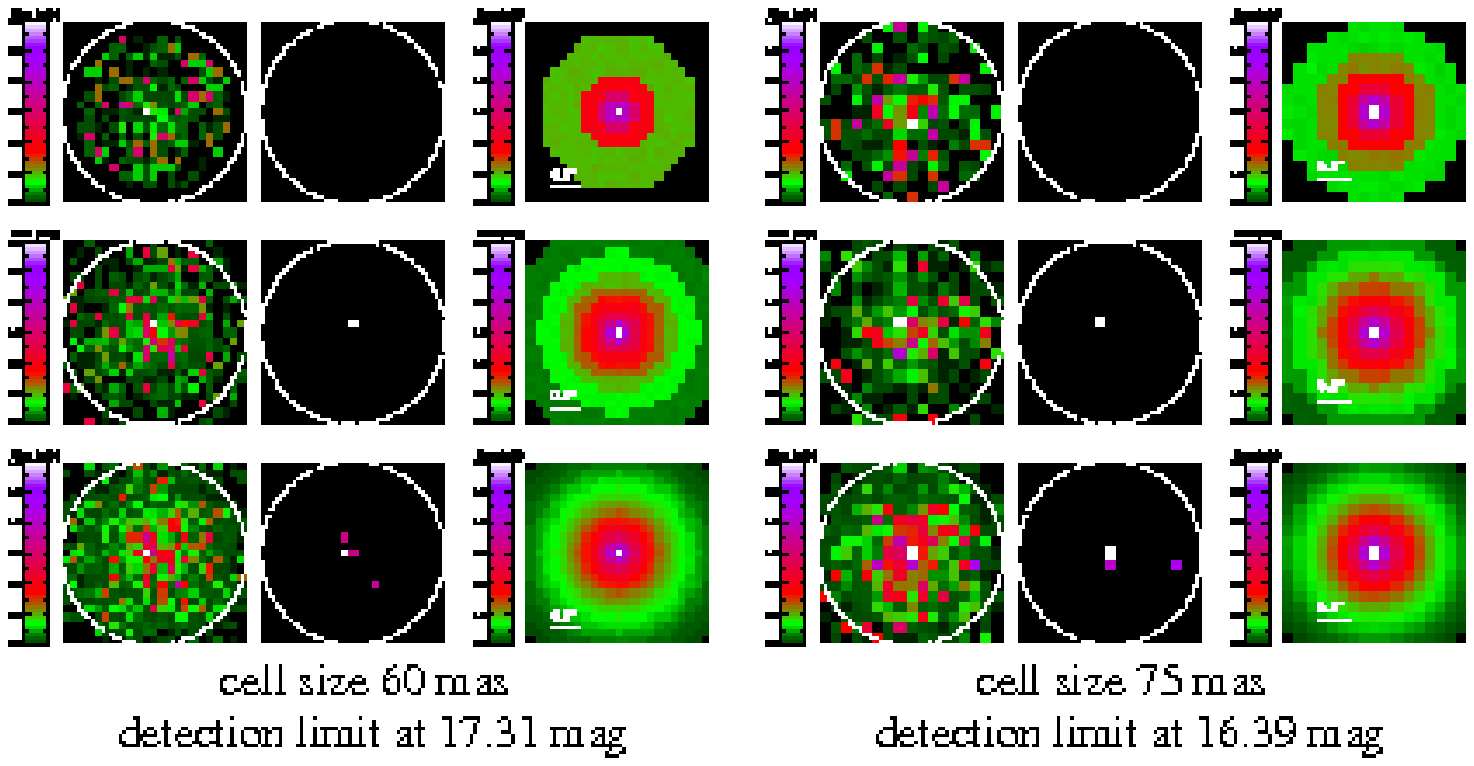}
  \caption{\small
Monte Carlo simulations for two different camera setups that resemble the
K'-band for NACO (right) and SINFONI (left) at the VLT.
{\bf Upper panel:} Snapshot of the simulation for the 0.11 KLF slope, power-law
index $\Gamma=0.3$ and 24.67 Ks-magnitude cutoff
{\bf Left:} All stars visible.
{\bf Middle:} Only the detectable blend stars visible. 
{\bf Right:} Average of all
the 10$^4$ simulation snapshots for the same setup.
Middle \& Lower panel: Same as top but for the 0.18 and 0.25 KLF slopes,
respectively.
}
  \label{eckartfig4}   
\end{figure*}

The simulations of the faint S-star cluster population was 
done by making use of the extrapolated KLF and star counts that go to
K-magnitudes fainter than 18 (with a cutoff at 25).
Sabha et al. (2012) distributed the stars in a $23 \times 23$ grid 
corresponding to 529 cells with
dimensions of $0.06'' \times 0.06''$. The cell size corresponds to
about one angular resolution element in K-band. 
So the simulations covers the inner $1.38'' \times 1.38''$ 
projected region surrounding Sgr~A*. 
Sabha et al. (2012) match the radial number density counts 
to the observed power-law index of $\Gamma = 0.30 \pm 0.05$ 
from Sch\"odel et al. (2007). 
Their algorithm  then fills each cell with its 
specified number of stars by choosing them randomly from a pool 
of stars created from the extrapolated KLF. 
Then the fluxes of the stars in each individual cell are added up
and compared to the flux of the faintest star detected in that region until now.
The simulations were carried out $10^4$~times 
in order to get reliable statistical estimates for the brightnesses 
in each resolution cell. 
Some of the results are depicted in Figure~\ref{eckartfig4}. 
The corresponding tables are given in  Sabha et al. (2012) for the
0.06~arcsec cell size and here in  
Tabs.~\ref{montecarlo} and \ref{montecarlocentral}
for the 0.075~cell size.
Sabha et al. (2012) 
find that especially towards the center - i.e. towards the
position of SgrA* - this phenomenon occurs for a quarter to 
three quarters of cases simulated.
This finding is consistent with the observations
(Witzel et al. 2012, Dodds-Eden et al. 2011) that find offsets to the
flux density towards SgrA* that are likely due to a time variable stellar 
contribution.
Using the observed KLF and number density slopes the total number 
of stars   in the simulated S-star cluster is consistent with the number 
of main sequence stars assumed in simulations by  Freitag et al. (2006).

With these simulations Sabha et al. (2012) could show 
that the formation of blend stars has to be considered. 
These will contaminate the appearance of the
S-star cluster at flux levels close to the confusion limit.
This is especially important at 
positions close to Sgr~A* where the number density counts peak.
The authors also point out that due to the sky-projected 
velocity dispersion of the stars within the S-star cluster 
of about 600~km/s, these blend stars will last for 
about 3-4 years before the apparent line of sight  clustering
dissolves and the blend star fades away. 
So it requires proper motion measurements 
over a time significantly longer than 3 years to decide if 
an object is in fact a blend star or a single object
for which reliable orbital parameters can be derived.
Spectroscopy may be difficult since the objects are faint.
Sabha et al. (2012) find that close to the center the probability 
of finding blend stars at any time is about 30\%-50\%. 
At the central position this effect may also give the
illusion of high proper motions.

The phenomenon of blend stars clearly demonstrate that higher angular 
resolution, astrometric accuracy and point source sensitivity
is required in order to further analyze the properties of 
the S-star cluster. 
In return these investigations  also greatly 
improve 1) the derivation of the amount and the compactness
of the central mass, 2)  the determination of relativistic 
effects in the vicinity of Sagittarius~A*, and 3) the determination
of the abundance of e.g. massive stellar black holes in the central 
S-star cluster. 
\\
\\
$\bullet$
{\it 
Finding stars fainter than those that are known in order 
to probe the gravitational potential requires an angular resolution (at high sensitivity)
beyond that achievable with current 8-10m telescopes.
This will be possible with the interferometers LINC-NIRVANA and GRAVITY 
in the NIR or with METIS at the E-ELT in the MIR.
}

\begin{figure*}
  \centering
  \includegraphics[width=15cm,angle=00]{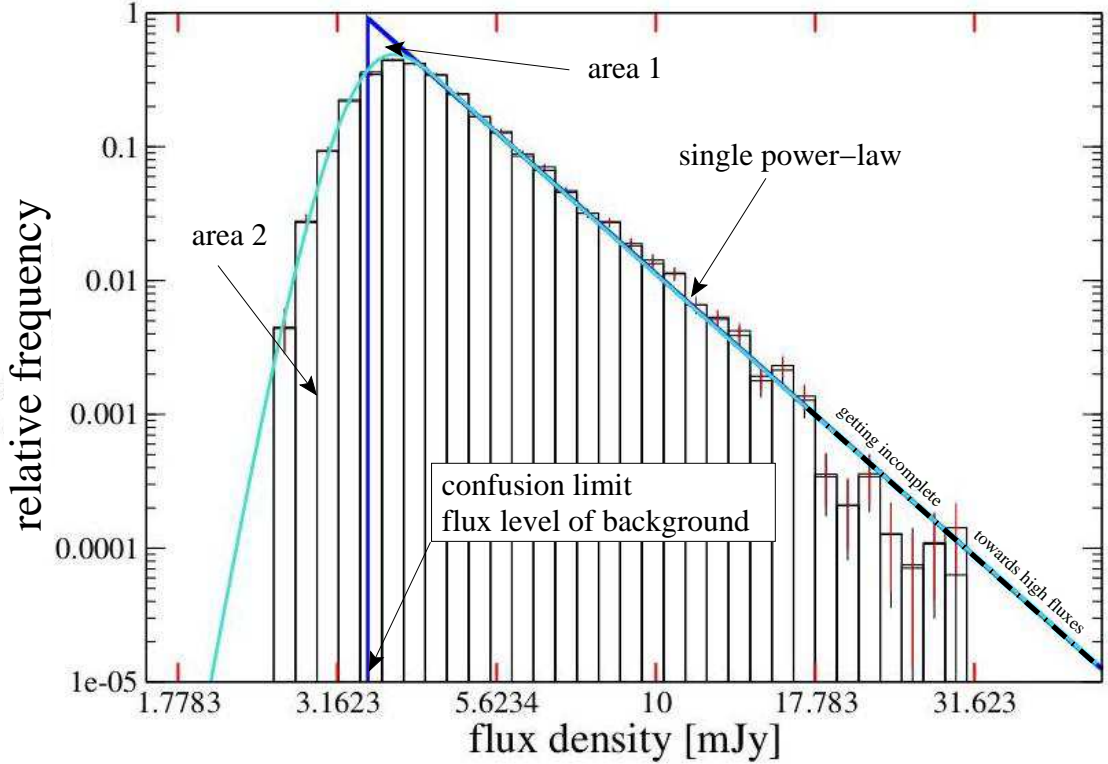}
  \caption{\small
Here we show the best piecewise-constant probability density model for
the flux densities of Sgr A* as presented by Witzel et al. (2012). 
The red error bars represent the uncertainty
of the bin height for the total amount of 10~639 data points.
In some cases there is a second bin height visible.
Those and the black error bars belong to the average histograms 
of 1000 datasets with a total of  6774 data points, 
generated by randomly removing points from the full dataset.
The dash-dotted blue line shows the extrapolation of the best power-law fit. 
The cyan line shows the power-law convolved with a Gaussian 
of a flux density distribution  of 0.32~mJy width (see Witzel et al. 2012).
The two areas labeled in the figure are equal in the corresponding 
non-logarithmic plot.
}
  \label{eckartfig1}   
\end{figure*}
\begin{figure*}
  \centering
  \includegraphics[width=15cm,angle=00]{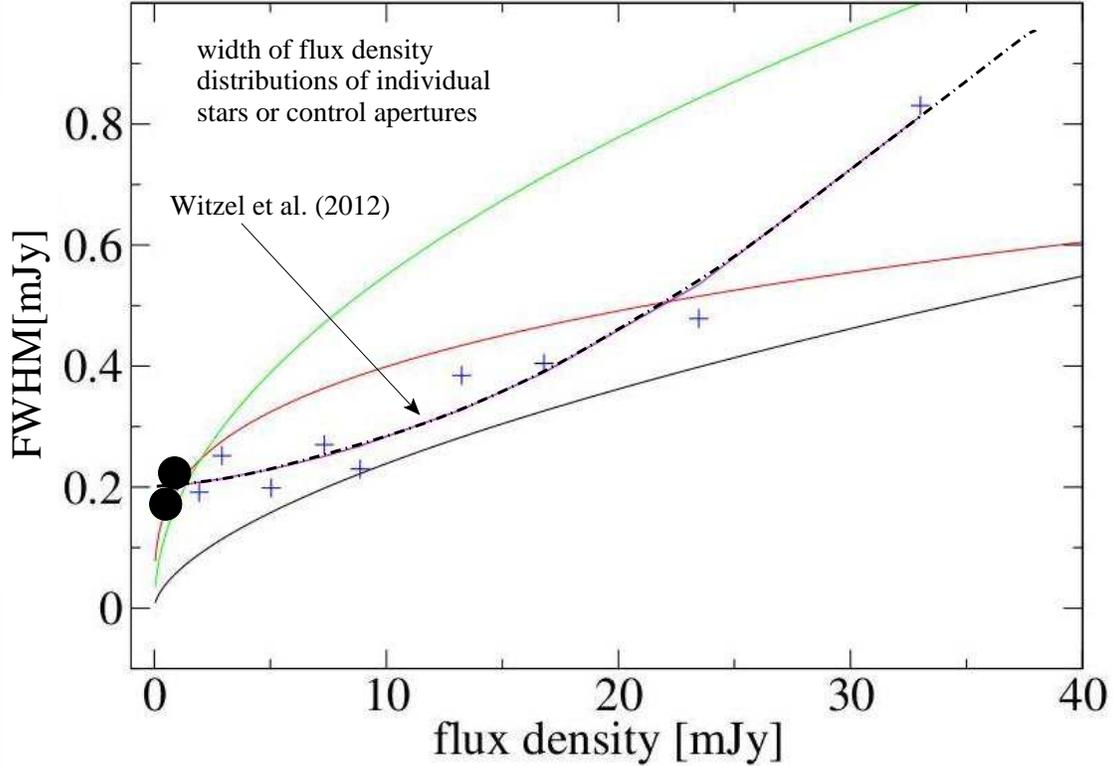}
  \caption{\small
The measurement error as a function of flux density
as presented by Witzel et al. (2012).
The blue line is a quadratic fit to the measured 
flux distributions of the calibration stars used to calibrate 
SgrA* flux densities as presented by Witzel et al. (2012).
The red line represents the power-law dependency given
by Do et al. (2009).
The green line and the black line show the dependency found by
Dodds-Eden et al. (2011) for their aperture photometry and their
PSF-fitting approach, respectively. 
The two black filled circles represent the measured widths of the 
flux density distributions at star-free control positions 
as described  by Witzel et al. (2012).
}
  \label{eckartfig2}   
\end{figure*}
\begin{figure*}
  \centering
  \includegraphics[width=15cm,angle=00]{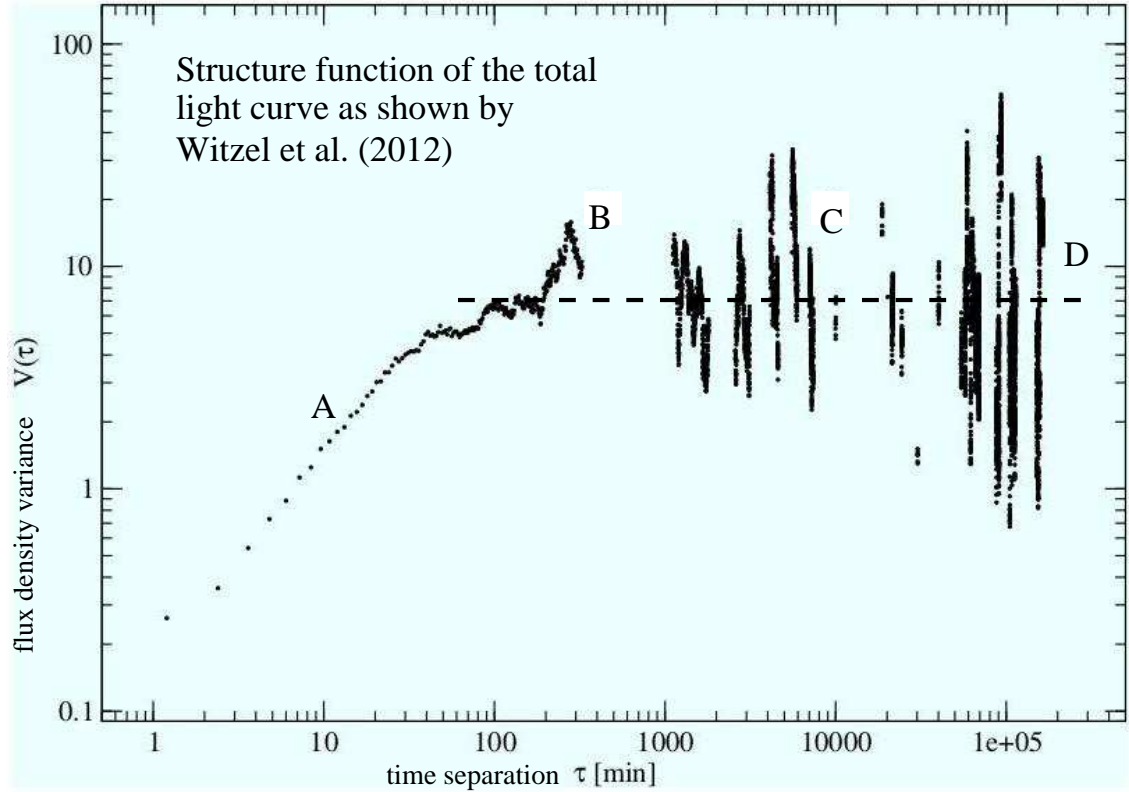}
  \caption{\small
  The structure function of the observed total light curve presented
by Witzel et al. (2012). The time binning is 1.2 minutes.
The flat region at large time separations has been indicated by a 
dashed line.
Specific regions have been labeled:
 A: around 20 minutes - domain 
were significant deviations from a straight line occurs;
B: around 1000 minute - day/night gap; 
C: around 10$^4$ minutes - time difference between observing runs within a year;
D: around few times 10$^5$ minutes - difference between summer/winter runs.
}
  \label{eckartfig3}   
\end{figure*}
\begin{figure*}
  \centering
  \includegraphics[width=12cm,angle=00]{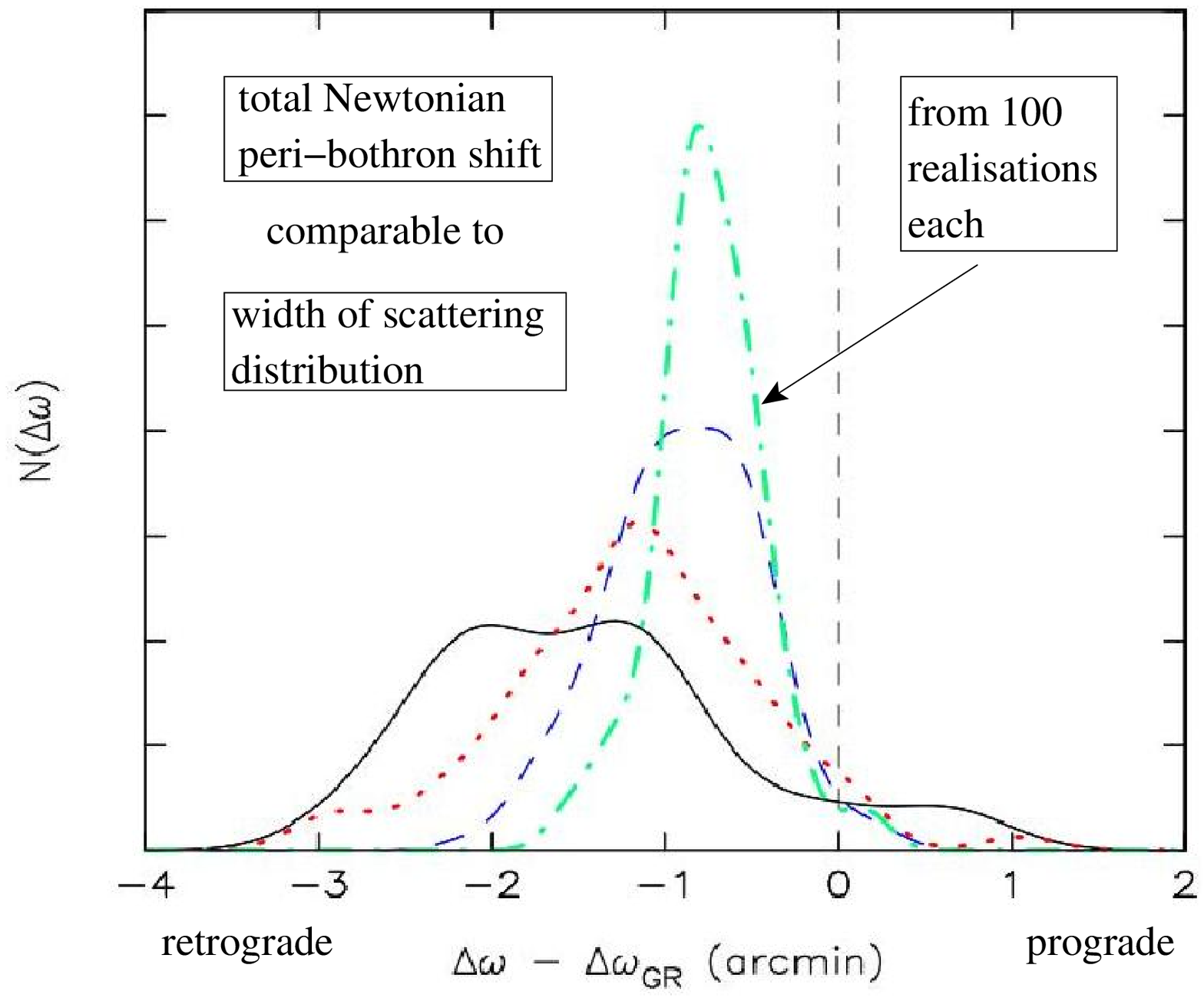}
  \caption{\small
Histograms of the predicted change in S2's argument of
peri-bothron, $\omega$, over the course one orbital period (about 16 yr).
The shift due to relativity, ($\Delta \omega$)$_{GR}$$\sim$11', 
has been subtracted from the total; 
what remains is due to Newtonian perturbations from the field stars. 
Regions of residual prograde and retrograde shift are indicated.
Each histogram was constructed from integrations
of 100 random realizations of the same initial model,
with field-star mass $m$ = 10~ \solm, and four different values of the
total number: $N$ = 200 (solid/black); $N$ = 100 (dotted/red);
$N$ = 50 (dashed/blue); and $N$ = 25 (dot-dashed/green). 
The average value of the peri-bothron shift increases with increasing
$Nm$. As discussed by Sabha et al. (2012) 
there is a separate contribution that scales approximately as
$\sim$1/$\sqrt(N)$
and that results in a variance about the mean value.
}
\label{eckartfig5}   
\end{figure*}

\subsection{Orbital elements of high velocity stars} 
\label{sec:orbits}

As laid out by Sabha et al. (2012), for faint stars close to the confusion 
limit the  determination of 
their position will be affected by the presence of background stars with
time variable clustering. 
The actual uncertainty in projected position, 
$\sigma_{\mathrm{position}}^2$,
will be affected by several effects.
Sabha et al. (2012) introduce $\sigma_{\mathrm{apparent}}^2$ as the 
apparent positional variation due to the photo-center variations of 
the star while it is moving across the projected distribution of 
fore- and background sources.
In addition to this there may an effect due to stellar scattering 
resulting in a variation of positions 
described by $\sigma_{\mathrm{scattering}}^2$. 
This may also contain a significant positional displacement due to the fact
that the stellar part of the gravitational potential may vary as a function 
of time due to the grainyness in the distribution of particularly the 
more massive stars. 
Finally, there is of course a systematic uncertainty 
$\sigma_{\mathrm{systematic}}^2$ that stems from
establishing and applying an astrometric reference frame.
Sabha et al. (2012) show how the different contributions can be 
derived or estimated such that information on 
$\sigma_{\mathrm{scattering}}^2$ can be obtained.
In addition of making use of data obtained from 8-10m telescopes,
one can alternatively make use of 
near-infrared interferometry with long baselines
and measure $\sigma_{\mathrm{scattering}}^2$ directly. 

Hence, measuring the time dependent position of S2 
interferometrically  with respect to brighter reference stars in the field,
may allow us to observe either single scattering events or
derivations from an elliptical orbit due to a varying underlying 
stellar fraction of the gravitational potential.
Sabha et al. (2012) point out that if scattering events 
contribute significantly to the uncertainties in 
the determination of the orbits, we will require  a larger number 
of stars to derive the physical
properties of the medium through which the stars are moving.
Averaging the results of $N$ stars that will then essentially sample
the shape of the distributions shown in Figure~\ref{eckartfig5}   
may therefore result in a $N^{-1/2}$ improvement 
in the determination of the extended mass.
At the same time one will obtain information on the composition of the
stellar environment the stars are  moving through.
\\
\\
$\bullet$
{\it 
Higher angular resolution observations as they will be possible with
LINC-NIRVANA,  GRAVITY in the NIR or METIS in the MIR
will help to overcome some of the 
confusion due to photo-center shifts caused by fore- and background stars.
Precise measurements of stellar trajectories are needed to 
distinguish between relativistic/Newtonian effects and effects due to 
resonant relaxation, i.e. stellar scattering or a variation of the 
stellar mass enclosed in the stellar orbits.}

\subsection{Faint states of Sagittarius~A*} 
\label{sec:faint}

Witzel et al. (2012) collected all Ks-band observations that had been 
carried out with the VLT until 2011. The authors constructed a 
double logarithmic histogram in which a given flux density 
value is plotted against its relative frequency.
The resulting flux density histogram is peaked and can be represented by a single 
power-law towards higher flux densities
(Figure~\ref{eckartfig1}).  
Witzel et al. (2012) conclude that it is not possible to verify the evidence 
of an intrinsic turnover that would indicate the
peaked shape of a log-normal distribution given the
knowledge about the true error distribution in the 
determination of flux densities at the
location of Sgr A* (see Figure~\ref{eckartfig2}).  
Therefore, having no evidence for a log-normal distribution 
for dim flux density states, the necessity of a break in the distribution to
account for the highest flux densities vanishes.
This is true even if one accepts the data selection by Dodds-Eden et al. (2011) and
ignores the fact that basically all flux density values higher than 17~mJy 
are only due to  a single bright event. 
Instead of a two state description of the flux density 
statistics of SgrA* Witzel et al. (2012)
prefer a simple power law with a slope of $\alpha$ = 4.2$\pm$0.1 
and an intrinsic x-axis offset of 3.57$\pm$0.1 mJy. 
The instrumental effects on the photometry lead to a detection limit -
and since flux density is a positive quantity, this intrinsic power-law
of SgrA* may break naturally very close to 0~mJy. 
Hence, a single power-law distribution
describes the observable intrinsic flux densities well.
This assumption is simpler
and clearly needs less parameters than the broken distribution
proposed by Dodds-Eden et al. (2011). 
It can not be excluded that 
the intrinsic flux distribution of SgrA* shows
some structure at flux densities below the detection limit,
i.e. between 0~mJy and 3.57~mJy.
It may as well follow a log-normal distribution 
at very low flux densities.
However, the log-normal distribution used in the 
model of Dodds-Eden et al. (2011) and a supportive 
evidence for a break in the observed flux density distribution
can be ruled out by the improved investigation of Witzel et a. (2012).
\\
\\
$\bullet$
{\it In oder to study the faintest flux density variations
of SgrA* in the near-infrared it is required to overcome the
confusion from fore- and background stars by increasing the 
angular resolution. This will be provided both by LINC-NIRVANA and GRAVITY.
LINC-NIRVANA has the advantage of being an imaging interferometer 
allowing for prompt calibration and extraction of flux densities from 
images. A disadvantage is the beam-shape allowing for high angular resolution
(for the Galactic Center from Arizona) in east-west direction, only.
For GRAVITY, flux density variations at high time resolution 
needs to be extracted from the visibilies directly or from 
imaging/modeling with variable uv-plane coverage.
In the MIR domain the high point source sensitivity of METIS at 
the E-ELT will allow to study faint flux density
variations against the stellar and thermal confusion background.
}

\subsection{High frequency quasi-periodicities} 
\label{sec:highfreq}
In Figure~\ref{eckartfig3}   we show the structure function of the 
entire light curve as investigated by Witzel et al. (2012).
Following the typical shape of a structure function 
there is a transition from a high frequency white noise 
regime via a straight line passing a turnover 
at a characteristic time scale to a flat regime.
We note that the first significant deviation from that straight 
line occures at a time scale of around 20 minutes. 
This is the same time scale that has previousely been 
pointed out as the time scale on which quasi periodic 
variations have been reported to happen on some occasions.
While any firm statement on that time scale based on the
structure function is hampered due to possible confusion from the
window function, this time scale is supported by statistics and individual
modeling of polarized flares, i.e. strong flux density excursions
(Eckart et al. 2006, Meyer et al. 2006a,b, 2007;
Zamaninasab et al. 2008, 2010, 2011).
\\
\\
$\bullet$
{\it For ground based light curves confusion from the time
windowing function cannot be overcome, unless very long
continous light curves from the southern hemisphere at locations
where SgrA* is circumpolar can be obtained. Alternatively 
flare statistics and modeling of individual flares in polarized light
have to be employed to further investigate the signatures of 
relativistically orbiting matter close to the central black hole 
(within or close to the mid-plane seen in GRMHD simulations).
Photo-center shifts as they will be measurable with GRAVITY 
as well as sub-mm VLBI observations will be useful 
to investigate the physics associated with the variability on
short time scales.}

\subsection{Radio variability of Sagittarius~A*} 
\label{sec:Radio}

The accretion flow onto SgrA* is often modeled as a relativistic
magneto hydrodynamic flow (GRMHD) with a central mid-plane close
to SgrA* where the densities and magnetic field strengths are 
likely to be highest
(Moscibrodzka et al. 2009, 2011, Dexter et al. 2010, 2012).
These models assume that the emission arises in a 
luminous flow region with an outer radius of about 
200$\mu$as, whereas the mid-plane covers the central
few 10$\mu$as (i.e. a few Schwarzschild radii).
The image of this structure is severely scatter broadened at radio
wavelengths longward of 3~mm.
Radio variability may and will occure from individual
regions (individual source components) all over the extended accretion flow.
At mm- and sub-mm-wavelengths this variability will
provide a significant confusion background against which the
central accretion phenomena close or within the mid-plane
has to be detected.
Characteristic frequency behaviours like adiabatic expansion
of source components can be clearly distinguished from
that confusion by following them over a frequency range that
is beyond the bandwidth of individual receiving channels.
A detailed discussion of this problem is given in Eckart et al. (2012).
\\
\\
$\bullet$
{\it Sub-mm observations over the largest possible
frequency range (as they are possible with ALMA) 
will help to distinguish mid-plane intrinsic
variability phenomena from the confusion originating within
the more extended accretion flow}

\subsection{X-ray variability of Sagittarius~A*} 
\label{sec:Xray}

In the X-ray domain it is difficult to study 
weak flux density variations. For faint states the flux density 
is dominated by the thermal Bremsstrahlung component with an
angular diameter of about one arcsecond (Baganoff et al. 2001, 2003).
Studying fainter flux density levels will allow an improved 
investigation of the detailed
correlation between the NIR and X-ray light curves.
\\
\\
$\bullet$
{\it Future missions with sub-arcsecond angular resolution and high 
sensitivity in the X-ray domain will allow to distinguish 
low flux density variations of SgrA* against confusion from 
the extended thermal Bremsstrahlung background component.}

\section{IMPLICATIONS FOR INSTRUMENTATION} 

 The described confusion effects clearly demonstrate the necessity of higher angular 
 resolution and point source sensitivity for future investigations of
 the S-star cluster and the derivation of the amount and the compactness
 of the central mass as well as the determination of relativistic 
 effects in the vicinity of Sgr~A*.
In the NIR/MIR domain these implications are clearly met by the upcomming
systems GRAVITY at the VLTI, LINC-NIRVANA at  the LBT and METIS at the E-ELT.

\acknowledgments     

\noindent
{\bf GRAVITY:}
Since 2008, this work is supported in parts by the German Federal
Department for Education and Research (BMBF) under the grants
Verbundforschung 05A08PK1 and 05A11PK2.
\\
{\bf LINC:}
Since 2001, this work is supported in parts by the German Federal
Department for Education and Research (BMBF) under the grants
Verbundforschung 05AL2PKA/5, 05AL5PKA/0 and  05A08PKA
\\
{\bf METIS:}
Since 2009, this work is supported in parts by the German Federal
Department for Education and Research (BMBF) under the grants
Verbundforschung 05A09PK1 and 05A11PK1. 
\\
\\
N.~Sabha is member of the Bonn Cologne Graduate School (BCGS) for
Physics and Astronomy supported by the Deutsche Forschungsgemeinschaft.
M.~Valencia-S. B.~Shahzamanian  and S. Smajic are members of the 
International Max-Planck Research School (IMPRS) 
for Astronomy and Astrophysics at the Universities of Bonn and Cologne
supported by the Max Planck Society.
Part of this work was supported by the German Deutsche 
Forschungsgemeinschaft, DFG, via grant SFB 956 and fruitful discussions 
with members of the European Union funded COST Action MP0905: 
Black Holes in a violent Universe and PECS project No. 98040.
M.~Garc\'{\i}a-Mar\'{\i}n and S. Fischer are supported by the German 
federal department for education and research (BMBF) under 
the project number 50OS1101.


\footnotesize
\vspace*{0.5cm}
\rf{Baganoff, F. K., Bautz, M. W., Brandt, W. N., et al. 2001, Nature, 413, 45}
\rf{Baganoff, F. K., Maeda, Y., Morris, M., et al. 2003, ApJ, 591, 891}
\rf{Brandl, B.R.; Lenzen, R.; Pantin, E.; Glasse, A.; et al.,  2010, SPIE 7735, 83}
\rf{Dexter, Jason; Fragile, P. Chris, 2012, 2012arXiv1204.4454D}
\rf{Dexter, Jason; Agol, Eric; Fragile, P. Chris; McKinney, Jonathan C. 2010, ApJ 717, 1092}
\rf{Do, T.; Ghez, A. M.; Morris, M. R.; Yelda, S.; Meyer, L.; Lu, J. R.; Hornstein, S. D.; Matthews, K., 2009, ApJ 691, 1021}
\rf{Dodds-Eden, K., Gillessen, S., Fritz, T. K., et al. 2011, ApJ, 728, 37}
\rf{Eckart, A. \& Genzel, R. 1996, Nature, 383, 415}
\rf{Eckart, A.; García-Marín, M.; Vogel, S. N.; Teuben, P.; Morris, M. R.; Baganoff, F.; Dexter, J.; Sch\"odel, R.; Witzel, G.; Valencia-S., M.; and 10 coauthors,  2012, A\&A 537, 52}
\rf{Eckart, A.; Schödel, R.; Meyer, L.; Trippe, S.; Ott, T.; Genzel, R., 2006, A\&A 455, 1}
\rf{Eisenhauer, F.; Perrin, G.; Brandner, W.; Straubmeier, C.; et  al.; 2011 Msngr, 143, 16}
\rf{Freitag, M., Amaro-Seoane, P., \& Kalogera, V. 2006, ApJ, 649, 91}
\rf{Genzel, R., Pichon, C., Eckart, A., Gerhard, O. E., \& Ott, T. 2000, MNRAS, 317, 348}
\rf{Ghez, A. M., Klein, B. L., Morris, M., \& Becklin, E. E. 1998, ApJ, 509, 678}
\rf{Gilmozzi, Roberto; Spyromilio, Jason, 2008, SPIE 7012, 43}
\rf{Herbst, T. M.; Ragazzoni, R.; Eckart, A.; Weigelt, G., 2010, SPIE 7734, 6}
\rf{Horrobin, M.; Eckart, A.; Lindhorst, B.; et al.,  2010, SPIE 7734, 58}
\rf{Morris, Mark, 1993, ApJ 408, 496}
\rf{Moscibrodzka, Monika; Gammie, Charles F.; Dolence, Joshua C.; Shiokawa, Hotaka; Leung, Po Kin, 2009, ApJ 706, 497}
\rf{Moscibrodzka, M.; Gammie, C. F.; Dolence, J.; Shiokawa, H.; Leung, P. K., 2011, ASPC 439, 358}
\rf{Muno, M. P.; Pfahl, E.; Baganoff, F. K.; et al. 2005, ApJ 622, L113}
\rf{Meyer, L., Eckart, A., Sch\"odel, R., et al. 2006a, A\&A, 460, 15}
\rf{Meyer, L., Sch\"odel, R., Eckart, A., et al. 2006b, A\&A, 458, L25}
\rf{Meyer, L., Sch\"odel, R., Eckart, A., et al. 2007, A\&A, 473, 707}
\rf{Sabha, N., Witzel, G., Eckart, A., et al. 2010, A\&A, 512, 13}
\rf{Sabha, N., Witzel, G., Eckart, A., et al. 2011, in Astronomical Society of the Pacific Conference Series, Vol. 439, Astronomical Society of the Pacific Conference Series, ed. M. R. Morris, Q. D. Wang, \& F. Yuan, 313}
\rf{Sabha, N., Eckart, A., et al., 2012, A\&A in press}
\rf{Schoedel, R., Eckart, A., Alexander, T., et al. 2007, A\&A, 469, 125}
\rf{Straubmeier, C.; Fischer, S.; Araujo-Hauck, C.; Wiest, M.; Yazici, S.; Eisenhauer, F.; Perrin, G.; Brandner, W.; Perraut, K.; Amorim, A.; Schoeller, M.; Eckart, A.; 2010, SPIE 7734, 97}
\rf{Vaitheeswaran, V.; Hinz, P.; O'Connell, C.; Kraus, J., 2010, SPIE 7740, 94}
\rf{Witzel, G., Eckart, A. et al. 2012, ApJ in press}
\rf{Zamaninasab, M., Eckart, A., Meyer, L., et al. 2008, Journal of Physics Conference Series, 131, 012008}
\rf{Zamaninasab, M., Eckart, A., Witzel, G., et al. 2010, A\&A, 510, 3}
\rf{Zamaninasab, M.; Eckart, A.; Dovciak, M.; Karas, V.; Sch\"odel, R.; et al., 2011, MNRAS 413, 322}

\end{document}